\documentstyle[12pt,epsfig]{article}
\begin{document}
\def\scri{\unitlength=1.00mm
\thinlines
\begin{picture}(3.5,2.5)(3,3.8)
\put(4.9,5.12){\makebox(0,0)[cc]{$\cal J$}}
\bezier{20}(6.27,5.87)(3.93,4.60)(4.23,5.73)
\end{picture}}

\noindent September 2008

\vspace{12mm}
 
\begin{center}

{\Large A NOTE ON TRAPPED SURFACES }

\vspace{10mm}

{\Large  IN THE VAIDYA SOLUTION}

\vspace{14mm}

{\large Ingemar Bengtsson}\footnote{Email address: 
ingemar@physto.se. Supported by VR.}

\vspace{5mm}

{\sl Stockholms Universitet, Fysikum\\
AlbaNova\\
S-106 91 Stockholm, Sweden}

\vspace{1cm}

{\large Jos\'e M. M. Senovilla}\footnote{Email address: josemm.senovilla@ehu.es. 
Supported by grants FIS2004-01626 (MEC) and GIU06/37 (UPV/EHU)}

\vspace{5mm}

{\sl F\'isica Te\'orica, Universidad del Pa\'is Vasco\\
Apartado 644\\
48080 Bilbao, Spain}

\vspace{14mm}

{\bf Abstract}

\end{center}

\noindent The Vaidya solution describes the gravitational collapse of a 
finite shell of incoherent radiation falling into flat spacetime and giving rise 
to a Schwarzschild black hole. There has been a question whether closed 
trapped surfaces can extend into the flat region (whereas closed outer 
trapped surfaces certainly can). For the special case 
of self-similar collapse we show that the answer is yes, if and only if 
the mass function rises fast enough. 

\newpage

{\bf 1. Introduction}

\vspace{5mm} 

\noindent What is a black hole? A possible answer is that a black hole is 
a region of an asymptotically flat spacetime bounded by an event horizon. 
Because the event horizon has a teleological character---it can be located 
only when the full solution is known---other answers have been sought 
\cite{Hayward, Ashtekar, Booth}. They rely on the occurence of trapped 
surfaces in the interior. A possible ``boundary'' of the black hole region 
might be a dynamical horizon, a spacelike hypersurface foliated by marginally trapped 
surfaces and typically lying well inside the event horizon. But there are 
difficulties with such a definition too. Here it seems pertinent to ask for the 
boundary of the region of spacetime through which closed trapped surfaces pass. 
It was conjectured by Eardley that a closed outer trapped surface passes 
through every point interior to the event horizon \cite{Eardley}, but if the 
surfaces are required to be genuinely trapped the corresponding statement is 
not true in general \cite{BenDov}. The boundary we ask for will lie somewhere 
in between a dynamical horizon and the event horizon (unless these coincide, 
as they do in the stationary case).  

In this note we address a very concrete question. Consider the spherically 
symmetric Vaidya solution \cite{Vaidya, Vaidya2}

\begin{equation} ds^2 = - \left( 1 - \frac{2m(v)}{r}\right)dv^2 + 2dvdr 
+ r^2d\theta^2 + r^2\sin^2{\theta}d\phi^2 \ . \end{equation}

\noindent Einstein's equations read 

\begin{equation} G_{ab} = 8\pi T_{ab} = \frac{2\dot{m}}{r^2}l_al_b \ , 
\hspace{10mm}  l_a = - \nabla_a v \ . \end{equation}

\noindent Thus $l_a$ is a null vector field, and the energy-momentum tensor is 
that of a radiation fluid. The energy conditions are obeyed provided that 
$\dot{m} \geq 0$, where the dot denotes differentiation with respect to $v$. But 
otherwise the rate at which radiation comes in is at our disposal. A simple choice, 
leading to a self-similar spacetime, is \cite{HWE}

\begin{equation} m = \left\{\begin{array}{cll} 0 & , & v \leq 0 \\ \mu v & , & 0 \leq v \leq M/\mu \\ 
M & , & v \geq M/\mu \end{array} \right. \ . \label{mass} \end{equation}

\noindent  This describes a finite shell of radiation entering flat spacetime from past 
null infinity, ending in a vacuum Schwarzschild black hole when the inflow 
stops. 

For the Vaidya solution Ben-Dov has shown that Eardley's conjecture is true \cite{BenDov}. 
However, this result concerns outer trapped surfaces, that is closed spacelike 2-surfaces 
whose outer null expansion is negative. For trapped surfaces, having both null expansions 
negative, Ben-Dov was able to show that there is a region of flat spacetime, inside the 
event horizon, into which no closed trapped surface can extend. The question whether closed 
trapped surfaces are able to extend into any part of the flat region was left open, and 
numerical investigations have so far not settled it \cite{Krishnan}. We will now proceed 
to answer this question for the special case of self-similar collapse. In a companion 
paper we discuss the issues in greater generality \cite{BS}.

\hspace{5mm}

{\bf 2. Basic facts}

\hspace{5mm}

\begin{figure}[!h]
\begin{center}
        \centerline{ \hbox{
                \epsfig{figure=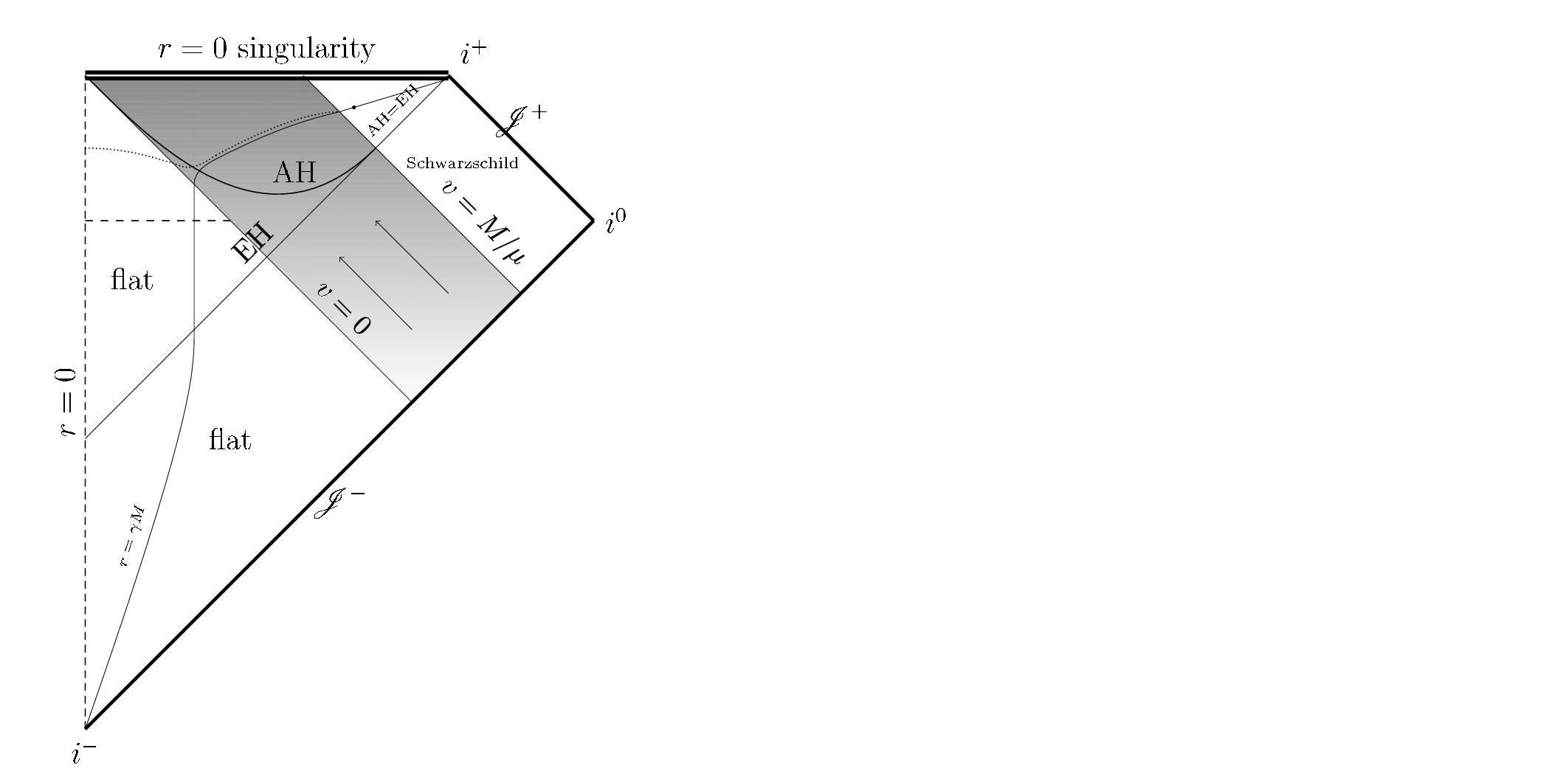,width=65mm}}}
\end{center}
        \caption{\small A Penrose diagram of the Vaidya solution for a mass function 
rising quickly enough for a closed trapped surface to penetrate into the flat region. 
It is shown as a dotted line emerging into the Schwarzschild part as a 
finite part of a line of constant $r$. Closed trapped surfaces cannot extend below 
the dashed line \cite{BS}. The part of flat space that lies inside 
the event horizon is shown without conformal distortion.}
        \label{fig:trap4}
\end{figure}

\noindent The solution is shown as a Penrose diagram in Fig. \ref{fig:trap4} (where it 
has been decorated with some extra structures to be encountered as we go on). 
Note that the part of flat space lying inside the black hole is a causal diamond: 
it is the intersection of the interior of two light cones, the event horizon and the 
inner part of the dust shell. The proper length of the geodesic connecting their 
vertices represents the maximum time an observer can live in this region, and will 
be calculated in eq. (\ref{lifetime}). 

The curved region certainly contains closed trapped surfaces. The round spheres $r =$ 
constant, $v =$ constant are trapped if and only if 
$r < 2m(v)$. Hence there is an apparent horizon at 

\begin{equation} r = 2m \hspace{5mm} \Longleftrightarrow \hspace{5mm} x \equiv \frac{v}{r} 
= \frac{1}{2\mu} \ . \end{equation}

\noindent Note the introduction of the variable $x$, which is a convenient one to use 
in the self-similar Vaidya region. In the Vaidya region the apparent horizon is a 
spacelike hypersurface, and in fact a dynamical horizon. 
In the self-similar case it is intrinsically flat, 
and the trace of its extrinsic curvature is 

\begin{equation} K = \frac{1-8\mu }{8v\mu^{3/2}} \ . \end{equation}

\noindent We know that a naked singularity will occur if and only if $\mu \leq 1/16$ 
\cite{HWE,Papapetrou}. Intuitively, the infinite mass density that would occur in a flat 
background becomes clothed only if the mass function rises fast enough, and we will 
assume that it does. Thus we study genuine black hole 
spacetimes without naked singularities, and it is our aim to construct closed 
trapped surfaces entering the flat region. 

We will need the null expansions for some special 2-surfaces. We quote them in 
a form that applies to any choice of the mass function. First we consider the 
2-surfaces

\begin{equation} \phi = \varphi \hspace{7mm} \theta = \frac{\pi}{2} \hspace{7mm} 
r = R(\rho ) \hspace{7mm} v = V(\rho) \ . \label{seven} \end{equation}

\noindent The first fundamental form is 

\begin{equation} d\gamma^2 = \Delta d\rho^2 + R^2d\varphi^2 \ , \hspace{8mm} 
\Delta \equiv V^\prime \left[ 2R^\prime - \left( 1 - \frac{2m}{R}\right) V^\prime 
\right] \ , \label{Delta} \end{equation}

\noindent where the prime stands for differentiation with respect to $\rho$. For 
spacelike 2-surfaces we demand $\Delta > 0 $, hence we must have $V^\prime \neq 0$. 
It follows that $R^\prime = 0$ can occur only inside the apparent horizon. 
Normal vectors to the 2-surfaces include 

\begin{equation} n = \frac{1}{\sqrt{\Delta}}(-R^\prime dv + V^\prime dr) \ , 
\hspace{5mm} e = Rd\theta \ ; \hspace{8mm} - n^2 = e^2 = 1 \ , \hspace{4mm} 
n\cdot e = 0 \ . \end{equation}

\noindent The corresponding null normals are 

\begin{equation} k_\pm = n \pm e \ . \end{equation}   

\noindent The null expansions are 

\begin{equation} \theta_\pm = \frac{1}{2\sqrt{\Delta}}\left[ \frac{1}{\Delta}\left( R^\prime 
V^{\prime \prime} - V^\prime R^{\prime \prime} + \frac{m}{R^2}R^\prime V^{\prime 2} - 
\frac{\dot{m}}{R}V^{\prime 3}\right) - \frac{R^\prime }{R} + \frac{V^\prime}{R}
(1 - \frac{m}{R})\right] \label{expansion1} \end{equation} 

\noindent and they are equal to each other. 

We will also consider the 2-surfaces 

\begin{equation} \phi = \varphi \hspace{7mm} \theta = \Theta (\rho) \hspace{7mm} 
r = r_0 \hspace{7mm} v = V(\rho) \ . \label{bojteta} \end{equation}

\noindent The first fundamental form is 

\begin{equation} d\gamma^2 = \Delta d\rho^2 + r_0^2d\varphi^2 \ , \hspace{8mm} 
\Delta \equiv \left( \frac{2m}{r_0}-1\right)V^{\prime 2} + r_0^2\Theta^{\prime 2} \ , 
\end{equation}

\noindent and again we demand $\Delta > 0 $. We assume that we are inside 
the apparent horizon, in which case normal vectors to the 2-surfaces include 

\begin{equation} n = \frac{1}{\sqrt{ \frac{2m}{r}-1}}dr \hspace{10mm} 
e = \frac{r\sqrt{\frac{2m}{r}-1}}{\sqrt{\Delta}}\left( V^\prime d\theta - \Theta^\prime dv - 
\frac{\Theta^\prime}{\frac{2m}{r}-1}dr \right) \ . \end{equation}

\noindent Again the null normals are given by $k_\pm = n \pm e$. The null expansions are 

\begin{eqnarray} \theta_\pm = \frac{\sqrt{\frac{2m}{r}-1}}{2\Delta^{\frac{3}{2}}} \times  
\hspace{72mm} \nonumber \\
\ \nonumber \\
\times \Big[ \pm r(\Theta^{\prime}V^{\prime \prime} - V^\prime \Theta^{\prime \prime}) 
+ \frac{\sqrt{\Delta}}{r}\left( (V^{\prime 2}(1 - \frac{m}{r}) - 2r^2\Theta^{\prime 2}
\right) \pm \frac{\Delta V^\prime}{r}\cot{\Theta} \Big] \label{expansion2} \\
\nonumber \\ 
- \frac{1}{\sqrt{\frac{2m}{r}-1}}\frac{1}{\Delta^{\frac{3}{2}}}(\sqrt{\Delta} 
\mp r\Theta^\prime )\frac{\dot{m}}{r}V^{\prime 2}  \ . 
\nonumber \end{eqnarray}
  
\noindent In the Schwarzschild region $\dot{m} = 0$, and the last term vanishes. 

\vspace{5mm}

{\bf 3. Plan}

\vspace{5mm}

\noindent The idea behind the construction to follow is simple. We start with a 
trapped surface in the flat region, say a hyperboloid ``bending down in time''. 
Topologically this is an open disk. It passes through the origin in flat space,
and meets the shell of radiation in 
a circle. In the Schwarzschild region as well we can find a trapped surface 
meeting the shell in a circle, say the cylindrical 2-surface 

\begin{equation} r = \mbox{constant} < M \ , \hspace{10mm} \theta = \frac{\pi}{2} 
\ . \end{equation}

\begin{figure}[!ht]
\begin{center}
        \centerline{ \hbox{
               \epsfig{figure=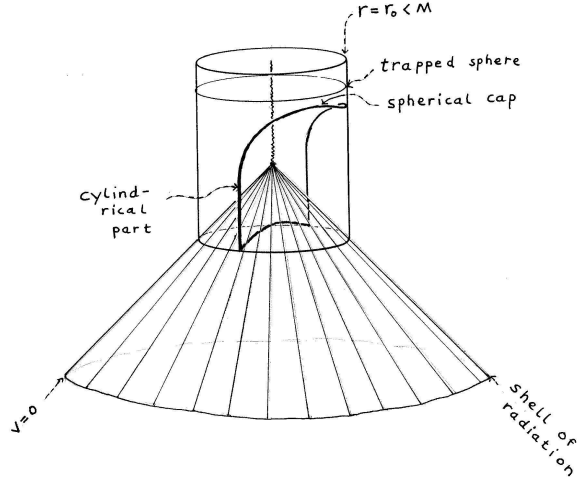,width=80mm}}}
\end{center}       
        \caption{\small A closed trapped surface we want to construct. The vertical coordinate is $t=v-r$, while the horizontal one is $r$. One dimension has 
been suppressed, and the 2-surface appears as a curve. The Vaidya region has been shrunk 
to a very thin shell, causing the trapped curve to 
have kinks as it passes through it.}
        \label{fig:trap1}
\end{figure}

\noindent By means of either eq. (\ref{expansion1}) or eq. (\ref{expansion2}) we can 
check that this cylinder does have both null expansions negative (and it will be 
marginally trapped if $r = M$). Now we must do two things: we must interpolate 
between these two surfaces in the Vaidya region, and we must ``cap'' the cylinder 
to ensure that the surface becomes closed. For the last part our idea is to join 
the cylinder to a hemispherical cap; whether this can be done while keeping both 
null expansions negative will be decided by eq. (\ref{expansion2}). Fig. \ref{fig:trap1} 
clarifies our plan. 
 
\vspace{5mm}

{\bf 4. A family of trapped surfaces in the flat and Vaidya region}

\vspace{5mm}

\noindent In flat space we obtain a marginally trapped (but not closed!) 
surface by setting $\theta = \pi/2$ and $t =$ constant, where $t = v - r$. We 
obtain a family of trapped surfaces of this type by choosing the hyperboloids 

\begin{equation} (t-t_0)^2 - r^2 = k^2 \ , \hspace{7mm} t = t_0 - 
\sqrt{r^2 + k^2} \ , \hspace{7mm} t_0 < k \ . \end{equation}

\noindent The expansions are 

\begin{equation} H(k_\pm) = - \frac{2}{k} \ . \end{equation}

\noindent In a $v-r$-diagram these surfaces obey 

\begin{equation} 0 < \frac{dv}{dr} < 1 \ . \end{equation}

\noindent In particular they meet the Vaidya region---at $v = 0$---with a slope 
that is less than one, but otherwise arbitrary. 

The Vaidya region requires more care. The surface will be defined by 
eqs. (\ref{seven}), with 

\begin{equation} R = \rho \ , \hspace{10mm} \frac{dV}{dR} = \frac{a}{b-X} \ , 
\hspace{5mm} X \equiv \frac{V}{R} \ .  \label{orddiff} \end{equation}

\noindent Here $a$ and $b$ are positive real numbers at our disposal. The differential 
equation can be solved straightforwardly \cite{Papapetrou, BS}, if we rewrite it as 

\begin{equation} R\frac{dX}{dR} = \frac{(X - \frac{b}{2})^2 + \Lambda^2}
{b-X} \ , \hspace{8mm} \Lambda^2 \equiv a - \frac{b^2}{4} \ . \label{papape} 
\end{equation}

\noindent To ensure that the solution has the properties we want it to, we 
choose 

\begin{equation} \Lambda^2 = a - \frac{b^2}{4} > 0 \hspace{5mm} \Leftrightarrow 
\hspace{5mm} 4\frac{a}{b} > b \ . \end{equation}

\noindent In particular this condition ensures that the first and second derivatives 
have the same sign, 

\begin{equation} \frac{d^2V}{dR^2} = \frac{1}{a^2R}\left( \frac{dV}{dR}\right)^3 
\left( \left( X-\frac{b}{2}\right)^2 + a - \frac{b^2}{4}\right) \ . \end{equation}

\begin{figure}[!t]
\begin{center}
        \centerline{ \hbox{
                \epsfig{figure=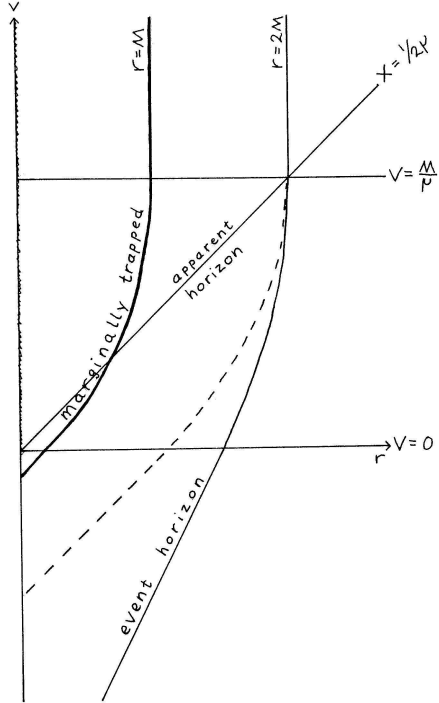,width=60mm}}}
\end{center}
        \caption{\small A $v-r$-diagram, drawn for the case $\mu = 1/2$. The trapped 
surfaces discussed in the text are represented by curves that become vertical at 
$v = M/\mu = 2M$, and lie to 
the left of the curve representing the marginally trapped surface. From the companion 
paper \cite{BS} we know for certain that no closed trapped surfaces can extend below the 
dashed line. Readers who prefer Penrose diagrams are referred to Fig. \ref{fig:trap4}.}
        \label{fig:trap2}
\end{figure}

\noindent In the $v-r$-diagram the curve becomes vertical at $X = b$. A look at 
Fig. \ref{fig:trap2} may be helpful here. 

We note in passing that radial null geodesics are given by the 
same differential equation (\ref{orddiff}), with $a = 1/\mu$, $b = 1/2\mu$. This allows 
us to locate the event horizon, and also to calculate the maximum time $\Delta t$ an 
observer can live in the flat region inside the black hole. This lifetime grows with 
$M$, and equals 

\begin{equation} \Delta t =4M\exp\left\{-\frac{2}{\sqrt{16\mu -1}}\arctan \frac{1}{\sqrt{16\mu -1}}\right\}<4M 
\ . \label{lifetime} \end{equation}

\noindent Thus, the larger the rate $\mu$ the sooner the event horizon starts to develop.  

The surface we have defined must be matched properly to trapped surfaces in the 
flat and Schwarzschild regions. To ensure that it matches to a trapped hyperboloid 
in the flat region we require 

\begin{equation} \left. \frac{dV}{dR}\right|_{X = 0} < 1 \hspace{5mm} \Rightarrow 
\hspace{5mm} \frac{a}{b} < 1 \ . \end{equation}

\noindent This ensures that the first derivative can be made continuous. There 
will be a finite jump in the second derivative. We will accept this, although in 
the end the construction can be made smoother, should one insist on it. The Schwarzschild 
region begins at $V = M/\mu$. At this boundary we want to join our surface to a trapped 
cylinder with $R = \gamma M$, where $\gamma < 1$. The curve is now vertical ($X = b$), 
so the requirement becomes 

\begin{equation} b = \frac{1}{\gamma \mu} \ . \end{equation}

\noindent Again this ensures that first derivatives are continuous. But now we 
have imposed a restriction on the mass function, namely 

\begin{equation} \mu = \frac{1}{\gamma b} > \frac{b}{4a \gamma} > \frac{1}{4\gamma} 
\ . \end{equation}

\noindent Recall that $b/a > 1$ and $\gamma < 1$. In the next section we will see that 
we must impose a somewhat stronger condition on $\gamma$ when we ``cap'' the surface 
in the Schwarzschild region. In the final section we observe that a restriction 
on $\mu$ of roughly this order can be deduced on general grounds, so the fact that 
such a restriction arises is not an artefact of the special way in which we construct 
the closed trapped surface.

We must ensure that our surface is spacelike. This will be so if $\Delta > 0$ in eq. 
(\ref{Delta}), and given $b > a$ this holds for all $X$ provided that 

\begin{equation} a \geq \frac{1}{\mu} \ . \end{equation} 

\noindent This is unproblematic. Finally we must check that the surface is indeed 
trapped. A calculation starting from eq. (\ref{expansion1}) shows that the null 
expansions are given by 

\begin{equation} \theta_\pm = - N\left[ (a\mu -1)\left( a + 3(b-a)X + (2a\mu - 1)X^2\right) 
+ (b-a)(2b-a)\right] \end{equation}

\noindent where 

\begin{equation} N = \frac{1}{2r\sqrt{a}}\frac{1}{(2b-a + 2(\mu a-1)X)^{3/2}} \ . 
\end{equation}

\noindent With the conditions already imposed the expansions are indeed negative, and 
we are done.

One additional detail is worth giving: Suppose the surface enters the Schwarzschild 
region at $r = r_S$. Then it enters the Vaidya region from flat space at 

\begin{equation} r_0 = r_S\ \mbox{exp}\left[ - \frac{b}{\Lambda}\arctan{\frac{b}{2\Lambda}}
\right] \ . \end{equation}

\noindent This follows \cite{BS} from the explicit solution of eq. (\ref{papape}).
 
To summarize, we have found a trapped surface that begins as a piece of a hyperboloid 
(say) in Minkowski space, continues through the Vaidya region, and joins the trapped 
cylinders at $r < M$ in the Schwarzschild region. It is depicted in Fig. \ref{fig:trap2}. But it 
is still not a closed surface. Two minor comments: The surface is only $C^{2-}$, but 
making it smoother presents no real difficulty. It is marginally trapped if 
$a = b = 1/\mu$, and if the hyperboloid is replaced by a plane in the flat region.

\vspace{5mm}

{\bf 5. Closing the trapped surfaces in the Schwarzschild region}

\vspace{5mm}

\begin{figure}[!t]
\begin{center}
        \centerline{ \hbox{
                \epsfig{figure=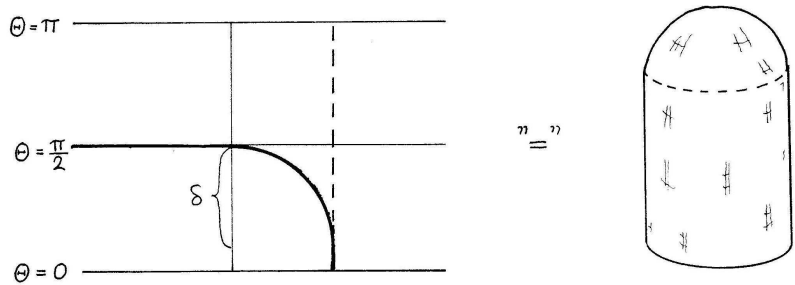,width=110mm}}}
\end{center}
        \caption{\small ``Capping'': In the $\Theta -V$-plane the straight line 
coming in from the left represents a cylinder, and the dashed line a round sphere. 
The curve interpolates between them with a quadrant of a circle, resulting in 
the intrinsic geometry shown on the right.}
        \label{fig:trap3}
\end{figure}

\noindent We are now safely in the Schwarzschild region, and our surface 
is joined to a cylinder at $r = \gamma M$, $\gamma < 1$. We must add a 
hemispherical cap to it. This piece of the surface will be of the form 
(\ref{bojteta}), with $V = \rho$. The surface is in effect described by a 
curve in the $\Theta -V$-plane, see Fig. \ref{fig:trap3}. The null expansions are given by 
eq. (\ref{expansion2}), and we can bring this to slightly more manageable form by 
introducing the dimensionless variable $\sigma$, 

\begin{equation} \sigma = V/\gamma M - \sigma_0 \ , \end{equation}  

\noindent where $\sigma_0$ is a constant we can adjust so that we start 
``bending'' the curve at $\sigma = 0$. Let $\dot{\Theta}$ be the derivative of 
$\Theta$ respect to $\sigma$, and assume that 

\begin{equation} \cot{\Theta} \geq 0 \ , \hspace{12mm} \ddot{\Theta} \leq 0 \ . \end{equation}

\noindent A calculation now shows that both null expansions will be negative if and 
only if   
 
\begin{equation} 2\dot{\Theta}^2 + \frac{1}{\gamma} - 1 > \sqrt{\frac{2}{\gamma} - 1 
+ \dot{\Theta}^2}\left( \cot{\Theta} - \frac{\ddot{\Theta}}{\frac{2}{\gamma}-1 + 
\dot{\Theta}^2}\right) \ . \label{trapping} \end{equation}

\noindent Our ``initial data'' are $\Theta = \pi/2$, $\dot{\Theta} = 0$ at $\sigma = 0$. 
We have not been able to handle the inequality in general. Let us just remark that as 
soon as a negative second derivative develops, there will be restrictions on $\gamma$ 
stronger than the condition $\gamma < 1$ that we have already imposed. 

A simple choice of the function $\Theta (\sigma)$, leading to a curve with continuous 
first derivatives, is to take a quadrant of the circle 

\begin{equation} \left( \Theta - \frac{\pi}{2} + \delta \right)^2 + \sigma^2 = \delta^2 
\ , \hspace{8mm} 0 < \delta \leq \frac{\pi}{2} \ .  \end{equation}

\noindent This joins the initial cylinder to a portion of one of the standard 
trapped round spheres at $\sigma = \mbox{constant} > 0$. At $\sigma = 0$ the trapping 
condition (\ref{trapping}) is obeyed provided that 

\begin{equation} \sqrt{\frac{2}{\gamma} -1}\left( \frac{1}{\gamma} - 1\right) 
> \frac{1}{\delta} \ . \end{equation}

\noindent Hence 

\begin{equation} \delta = \frac{\pi}{2} \hspace{5mm} \Rightarrow \hspace{5mm} 
\gamma < 0.68514 \ . \end{equation}

\noindent We have checked numerically that with this restriction the inequality 
(\ref{trapping}) is obeyed for all $0 \leq \sigma \leq \pi/2$, so this does it: we 
do have a closed trapped surface that begins in the flat region and ends in 
Schwarzschild. Topologically it is a 2-sphere.

Clearly the construction can be improved by changing the precise shape of the 
curve in the $\Theta - V$-plane. The best we were able to do (using a segment 
of an ellipse rather than a segment of a circle) was $\gamma < 0.75$. Another 
possibility would perhaps be to start ``capping'' the surface already in the 
Vaidya region, but we have not considered this.

\vspace{5mm}

{\bf 6. Discussion}

\vspace{5mm}

\noindent In the self-similar Vaidya solution we have constructed closed trapped 
surfaces that begin in the flat region, pass through the shell, and end in 
the Schwarzschild region. See Fig. \ref{fig:trap4}. 
The mass function is given by eq. (\ref{mass}) and the 
construction works provided that 

\begin{equation} \mu > \frac{1}{0.68} \frac{1}{4} \ . \label{result1} \end{equation}

\noindent On the other hand we know from the companion paper \cite{BS}, using 
an argument based on the interplay between the Kerr-Schild vector field 
\cite{Hildebrand} and the trapped surfaces, that it is impossible to construct 
a closed trapped surface entering the flat region if 

\begin{equation} \mu \leq \frac{1}{8} \ . \label{result2} \end{equation}

\noindent It will be observed that there is a gap between these two results.  

Indeed our construction is not optimal. We indicated a way in which the 
number $0.68$ in inequality (\ref{result1}) can be increased somewhat. It must 
however be kept smaller than one for the kind of surface we consider. We do not 
know what one can achieve if one gives up axial symmetry of the surfaces, say. 
On the other hand we do know that the inequality (\ref{result2}), which rules 
out closed trapped surfaces extending into the flat region, can be relaxed. The 
true limit must lie somewhere in between. 

Like Ben-Dov's construction of closed outer trapped surfaces \cite{BenDov}, 
our result serves to underline the fact that the teleological character of event horizons 
has a translation into a ``non-local'' property of closed trapped surfaces.
They can live partly in a region of spacetime whose entire past is flat, if energy falls 
across them elsewhere to make their closure possible.

\vspace{1cm}

{\bf Acknowledgements}

\vspace{5mm}

\noindent We thank Jan \AA man for performing some checks. Jos\'e M.M. Senovilla 
thanks the Physics Department at Stockholm University for its hospitality; 
we both thank the Wenner-Gren Foundation for making his visit possible.


\begin{thebibliography}{99}

\bibitem{Hayward} S. A. Hayward, {\it Black holes: New horizons}, in V. G. Gurzadyan 
et al. (eds.): Proc. of the Ninth Marcel Grossmann meeting, World Scientific 2002.

\bibitem{Ashtekar} A. Ashtekar and B. Krishnan, {\it Isolated and dynamical 
horizons and their applications}, Living Rev. Rel. {\bf 7}, 10 (2004).

\bibitem{Booth} I. Booth, {\it Black hole boundaries}, Can. J. Phys. {\bf 83} (2005) 
1073.


\bibitem{Eardley} D. M. Eardley, {\it Black hole boundary conditions and coordinate 
conditions}, Phys. Rev. {\bf D57} (1998) 2299.

\bibitem{BenDov} I. Ben-Dov, {\it Outer trapped surfaces in Vaidya spacetime}, 
Phys. Rev. {\bf D75} (2007) 064007. 

\bibitem{Vaidya} P. C. Vaidya, {\it The gravitational field of a radiating 
star}, Proc. Indian Acad. Sci. {\bf A33} (1951) 264.

\bibitem{Vaidya2} P. C. Vaidya, {\it 'Newtonian' time in general relativity}, 
Nature {\bf 171} (1953) 260.

\bibitem{HWE} W. A. Hiscock, L. G. Williams and D. M. Eardley, {\it Creation of particles 
by shell-focusing singularities} Phys. Rev. {\bf D26} (1982) 751

\bibitem{Krishnan} E. Schnetter and B. Krishnan, {\it Non-symmetric trapped 
surfaces in the Schwarzschild and Vaidya spacetimes}, Phys. Rev. {\bf D73} 
(2006) 021502(R).

\bibitem{BS} I. Bengtsson and J. M. M. Senovilla, {\it The boundary of the 
region with trapped surfaces in spherical symmetry}, to appear.

\bibitem{Papapetrou} A. Papapetrou, {\it Formation of a singularity and causality}, 
in N. Dadhich et al. (eds.): A Random Walk in Relativity and Cosmology, Wiley 1985.

\bibitem{Hildebrand} B. Coll, S. R. Hildebrandt and J. M. M. Senovilla, 
{\it Kerr-Schild symmetries}, Gen. Rel. Grav. {\bf 33} (2001) 649.

\end{thebibliography}
\end{document}